# Applying Titius-Bode's Law on Exoplanetry Systems

M. B. Altaie, Zahraa Yousef, A. I. Al-Sharif

Department of Physics, Yarmouk University, 21163 Irbid, Jordan

We report the application of Titius-Bode's law on 43 exoplanetary systems containing four or more planets. Due to the fact that most of these systems have their planets located within compact regions extending for less than the semi-major axis of Mercury we found the necessity to scale down the Titius-Bode law in each case. In this short article we present sample calculations for three systems out of the whole set. Results show that all systems studied are verifying the applicability of the law with high accuracy. Consequently our investigation verifies practically the scale invariance of Titius-Bode law. The results of this study buildup the confidence in predicting positions of the exoplanets according to Titius-Bode's law besides enabling diagnosing possible reasons of deviations.

# INTRODUCTION

During the last two decades many exoplanetary systems have been discovered, most of these contain one or two planets, but a large number of systems contain more than three exoplanets. It would be worth to study the spacing between these planets in order to know whether the spacing satisfies the so-called Titius-Bode (hereafter TB) law or not. A positive result will provide support for this law and may inspire more studies to find its physical foundation.

Historically the first support for the law came through its role in discovering Uranus where the position of the seventh planet in our solar system was predicted at 19.6 AU by this law. In 1781 William Herschel found that there is a planet at 19.218 AU which deviates only by 1.9% from the expected distance. Uranus could have been discovered earlier if the TB relation had been taken more seriously.[1] Although the TB law did not give an accurate prediction for the distance of Neptune from the Sun, it played an important role in discovering the Asteroid Belt, where the median of position was predicted with high accuracy. Many modified formulas were presented for the TB law because of the large deviation suffered in its prediction in the case of the outer planets. The first modification was suggested by Wurm in 1803 and the last to date was in 2009 presented by Pankovic and Radacovic.[2]

The configuration of the discovered exoplanetary systems can be compared with our solar system. In a given multiple exoplanetary system there could be planets that have not been discovered yet. That may occur because of the lack of sufficient data, such as planetary masses or radii that are below the detection limits.[3] Here the TB law could be a useful tool for making predictions about the positions of exoplanets in multiple exoplanetary systems. The aim of this study is to test this idea. We first try to check if the multiple exoplanetary systems fit well with the TB law, then we use the law to make predictions about the positions of undetected exoplanets in systems under our study.

Several authors have considered applying the TB law on spacing between of the exoplanets in multiple planetary systems. These studies followed different approaches and have different strategies. Poveda and Lara[4] tried to investigate if the five planets of the system 55 Cancri (also called 55Cnc) fit the TB law. They found that if the largest known semi-major axis to be allocated for the sixth planet then the exponential form of the TB law fits well with the five observed semi-major axes for that system. Hence they predict the existence of a planet at $a=0.2$ AU in the large gap between the fourth planet (at $a=0.781$ AU) and the sixth (at $a=5.22$ AU). Moreover, they predicted another new planet for the system at $a=15.0$ AU. After publishing their paper, the semi major axis for the most inner planet was updated from 0.038 AU to 0.015AU as reported by Winn.[5]

Raymond et al.[6] mapped out the region in 55 Cancri between the known planets 55Cnc f and 55Cnc d where an additional planet, designated g, might exist. Because there is a wide region of stability between them, they found that this region can accommodate a Saturn-mass planet which is a giant planet so they suggested that there could be even two or three additional planets.

Lovis et al.[7] applied the generalized TB relation to multiple exoplanetary systems detected by the radial velocity method only, they found reasonable fits. But they did not make any predictions for positions of new planets.

Cuntz[8] used the TB relation to predict the existence of new planets in the 55 Cancri system. He predicted the existence of four planets at distances at 0.085 AU, 0.41 AU, 1.50 AU and 2.95 AU from their host star. He also predicted the distance of the next possible outer planet in the system to be between 10.9 AU to 12.2 AU. None of the above predictions is confirmed observationally.

Bovaird and Lineweaver[9] (2013) studied the applicability of TB relation on a sample of multiple exoplanetary systems containing four planets and more. They found that the majority of their sample adheres to the TB relation more than the solar system does. They inferred that if any system does not follow the TB relation as close as our Solar System there is a high probability that one or more exoplanets in this system are not detected so far. They used the TB relation to list their predictions for the existence of 141 exoplanets in 68 multiple exoplanetary systems. Huang and Bakos (see ref. 3) used the Kepler mission data to search 97 planets of those studied by Bovaird and Lineweaver (ref. 9) in multi-planetary systems; they only found five planetary candidates around their predictions. They considered that the remaining predicted planets could not be discovered by Kepler mission yet. They conclude that the ability of the TB law at making predictions in extrasolar planetary systems is questionable. In this letter we present sample calculations for three exoplanetary systems and report a summary for our investigation of the 43 systems containing planets giving the results for the adherence to the TB law presented in terms of an average of the fitting percentage. Detailed calculations for the set of the 43 systems are to be given elsewhere.[10]

## Method and Results

The strategy of our study is based on the understanding that TB law is describing a configuration rather than a law with fixed numbers. This means that the numbers given by the sequence 0.4, 0.7…, etc. are only set as a scale for the configurations. The way that the TB law was deduced confirms this understanding. Accordingly, we feel that a scaling of the configuration is always possible, and therefore can be taken as a reference for the application of the TB law as long as the relative arrangement is preserved. From this point of view this study is a novel one that could open wide door for other studies in the field and motivating new constrains on the way exoplanetary systems are configured.

We prepared a survey for the exoplanets reported by NASA Exoplanets Archive about the multiple exoplanetary systems under study. The confirmed exoplanets discovered up to date (21/1/2016) are 1932 planets. 259 of these exoplanets belong to 58 multiple exoplanetary systems (in which the same star hosts more than 3 planets). We have excluded 15 multiple exoplanetary systems with 66 exoplanets for lack of data about their semi major-axes. Our sample contains the remaining 193 exoplanets that are hosted by 43 stars. Thirty five systems host 153 of these planets were detected by the transit method, seven systems involves 36 planets were detected by the radial velocity method and only one system (HR8799) with 4 planets was detected by direct imaging. Our data about exoplanets are obtained from the NASA Exoplanets Archive.[11]

## TB law scaling

We noticed that most of the known exoplanetary systems under study have their planets located within regions of distances less than the semi-major axis of Mercury. This means that we need to scale the TB law for each of those systems. For this purpose our method was to choose one of the planets of the given system to be satisfying the TB law at a given order of our choice, calculate the scaling factor which comes to be the ratio between the planet's semi-major axis and the corresponding distance in TB law, accordingly we go on scaling all the system with the same factor. Such a scaling of the distances may also be understood as a scaling of the measuring units as we can calculate this factor for any system by dividing the new astronomical unit by the corresponding distance in the TB law. Every individual system has its own scaling factor. The choice of the order of the planet to fit the TB law is done with the best fit and minimum deviation from the distance given by the law for that order. Once we get the scaling factor the configuration of the system becomes known to us. Accordingly, we can place the planets in their positions and the order of TB law which we have chosen up to 10 positions. This resulted, in many cases, in leaving some empty positions for planets that are not discovered yet, or that which even may not have been formed at all as it is the case with the Asteroid Belt. These empty positions are considered as predictions from our point of view.

The justification for our scaling was theoretically validated by Garner and Dubrulle[12] who had showed that the law is both scale and rotation invariant. This important work has established the theoretical bases for the invariance of TB law and it our work which we are presenting here is the practical verification for that theoretical work of Garner and Dubrulle, where the results shows that the scaled exoplanetary systems are satisfying the TB law with high accuracy, on average is 90.4%.

# Fitting results

Out of the 43 systems that we have studied, we will choose only three systems. The first is the one with the best fitting percentage of 97.78% which is Kepler 215 system. The second is arbitrarily chosen from the systems which showed near the average fitting percent of about 90.4% and the last system is the one which showed the worst fitting percentage of 65.27%, this is Kepler 444 system.

**KEPLER-215**

This is a G-type star similar to our Sun, its mass is about 0.98 solar mass with radius of about 1.0 solar radius and surface temperature of about 5739K. This exoplanetary system harbors 4 confirmed planets which were detected by the transit method. The positions of the confirmed planets in the system are as shown in the second column of Table (1). The scaling factor for this system was calculated by positioning the second planet which has a semi-major axis of 0.311 AU in order n=3, accordingly the scaling factor should be 0.113. The planets configuration becomes as shown in Table (1). We predict an inner planet at order n=1 positioned at 0.0452 AU and the rest of the predicted planets are all in the outer region at 0.5876 AU and beyond. We have shown the absolute and the relative fitting errors marking the deviation from the standard TB law in order to expose the error contribution from each planet. This might be useful on analyzing the pathology of the system and may help designating some reasons for deviations from the TB law. With the suggested configuration the average fitting percentage for the system is calculated to be 97.78%. If we look at the configuration we see high similarity with the solar system apart from the scale factor.

Table (1) System Kepler-215

| n | $a_o$ | $a_p$ | abs. error | rel. error |
|---|---|---|---|---|
| 1 | - | 0.0452 | - | - |
| 2 | 0.0840 | 0.0791 | 0.0049 | 0.0583 |
| 3 | 0.1130 | 0.1130 | 0.0000 | 0.0000 |
| 4 | 0.1850 | 0.1808 | 0.0042 | 0.0227 |
| 5 | 0.314 | 0.3164 | 0.0024 | 0.0076 |
| 6 | - | 0.5876 | - | - |
| 7 | - | 1.1300 | - | - |
| 8 | - | 2.2148 | - | - |
| 9 | - | 4.3844 | - | - |
| 10 | - | 8.7236 | - | - |
| Avg. rel. error | | | | 2.22% |
| Fitting percent | | | | 97.78 |

The fitting of the planets for this system is shown in Fig. (1).

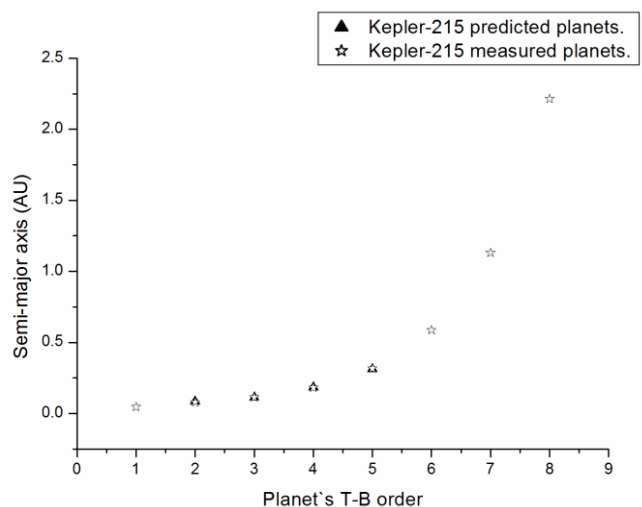

**Figure (1) The best TB fitting possible for system Kepler-215. This gives the highest fitting percentage among the set of exoplanetary systems considered in our study. The triangles mark the data from observations and the stars mark data from our predictions.**

**HD 160691**

The star is again a G3 type that has a mass of about 1 solar mass with radius of 1.245 solar radius. The surface temperature of the star is about 2700K. The age of the star is not well defined but there are some references claiming it is about 6 Gyrs old. The system contains four confirmed planets that have been discovered by radial velocity method. The semi-major axes of the confirmed planets are as shown in the second column of Table (2). Clearly the system has a very large gap between the 3$^{rd}$ and the 4$^{th}$ planet where the distance is about 3.738 AU, a large distance in comparison with the size of the system. Consequently, the fitting allow us to predict the existence of several planets: an inner one at 0.0540 AU, and three intermediate ones at 0.1349 AU, 0.2159 AU and at 0.3778 AU. Two more planets are predicted in the outer region at 1.497 AU and at 2.6445 AU. Outer planets at 10.416 AU and more may also exist. The average fitting percentage for this system is 90.61% which is within the average obtained for the 43 systems which we have studied.

Table (2) System HD 160691

| n | $a_o$ | $a_p$ | abs. error | rel. error |
|---|---|---|---|---|
| 1 | - | 0.0540 | - | - |
| 2 | 0.0909 | 0.0944 | 0.0049 | 0.0583 |
| 3 | - | 0.1349 | - | - |
| 4 | - | 0.2159 | - | - |
| 5 | - | 0.3778 | - | - |
| 6 | 0.9210 | 0.7016 | 0.2194 | 0.2382 |
| 7 | 1.497 | 1.3492 | 0.1478 | 0.0987 |
| 8 | - | 2.6445 | - | - |
| 9 | 5.2350 | 5.2350 | 0.0000 | 0.0000 |
| 10 | - | 10.416 | - | - |
| Avg. rel. error | | | | 9.39% |
| Fitting percent | | | | 90.61 |

The fitting is chosen to have the fourth planet in position n=9. This will make the scaling factor become 0.13492. Accordingly the configuration of the planets for the best fitting is obtained as gives in Table (2). The fitting curve for the system is as depicted in Fig. (2).

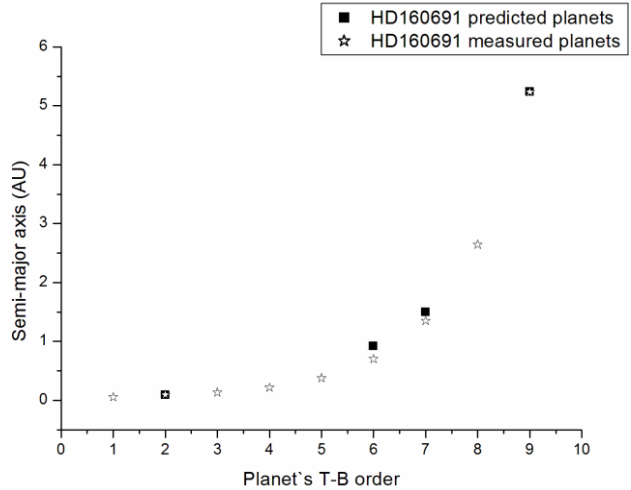

**Figure (2) The second best TB fitting possible for system HD 160691. This has been chosen for display because it has a fitting percentage of about the average for the set studied. The dark squares mark the data from observations and the stars mark data from our predictions.**

## KEPLER-444

This is a system with extreme dynamical conditions. It harbors three stars (triple star system). There are 5 confirmed planets orbiting Kepler 444A which is a K-type star in a tightly packed region. The mass of this primary star is about 0.76 solar mass with a radius of about 0.75 solar radius. This star is believed to be 11.2 Gyrs old. This is the oldest known star to date harboring exoplanets. The other two stars are Kepler 444B and Kepler 444C are small M-type stars orbiting the center of mass of the whole system at a distance of about 40 AU. So the whole known system composed of the three stars and the five planets has a size less than our solar system. As for the Kepler 444A system the five planets orbit the host star at distances shown in Table (3). All the predicted planets are in the outer region at 0.2262 AU and beyond.

Table (3) System Kepler-444

| n | $a_o$ | $a_p$ | abs. error | rel. error |
|---|---|---|---|---|
| 1 | 0.0418 | 0.0174 | 0.0244 | 0.5835 |
| 2 | 0.0488 | 0.0304 | 0.0148 | 0.3762 |
| 3 | 0.0600 | 0.0435 | 0.0165 | 0.2750 |
| 4 | 0.0696 | 0.0696 | 0.0000 | 0.0000 |
| 5 | 0.0811 | 0.1218 | 0.0407 | 0.5018 |
| 6 | - | 0.2262 | - | - |
| 7 | - | 0.4350 | - | - |
| 8 | - | 0.8526 | - | - |
| 9 | - | 1.6878 | - | - |
| 10 | - | 3.3582 | - | - |
| Avg. rel. error | | | | 34.73% |
| Fitting percent | | | | 65.27 |

To obtain a fitting for this system we found that the best possible fit will be obtained if we position the 4th planet in n=4 of the TB order. This gives a scaling factor of 0.04350. Consequently we obtain the configuration given in Table (3) where we see the confirmed planets are arranged sequentially. The fitting percentage is 65.27, which is very poor as compared to the other planetary systems in this study. The reason is clear; the dynamical setup of the system having three stars at the play makes the system highly unstable. The orbits of this system are coinciding. Orbits of planets c and d, e and f are coinciding. This causes instability in the system.
The fitting curve of the system is shown in Fig (3).

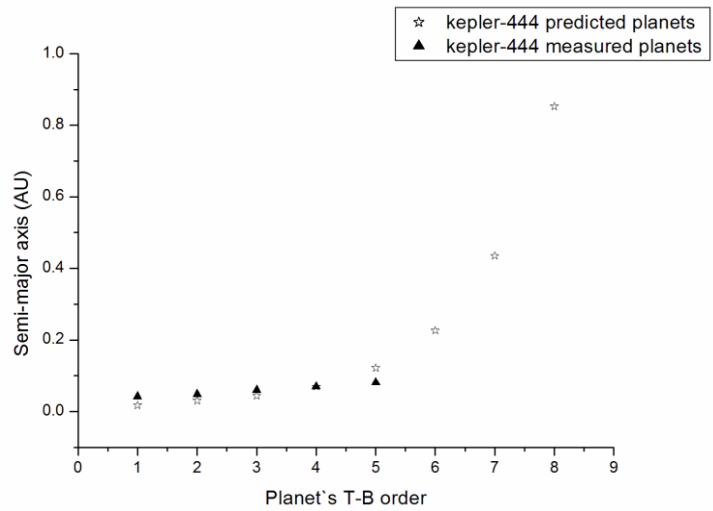

**Figure (3) The best fitting possible for system Kepler-444. This gives the lowest fitting percentage among the set of exoplanetary systems considered in our study.**

## Discussion and conclusions

In this letter we have chosen to present three systems out of the set of 43 exoplanetary systems which have been fitted in accordance with the TB law. We find that the majority of our large set adheres to the TB law with an average adherence of about 90.40%. The examples given in this letter have been chosen to show the highest, the median and the lowest of the fitting percentages. Generally we notice that the adherence to the TB law for exoplanetary systems under our study has different percentages. There are several reasons for these differences; one of them is the age of the system indicated by the age of the hosting star. Young systems with age of less than 1 Gyr are normally under formation and might not have dynamically settled yet. Such systems suffer jostling during planetary system formation; this is a messy process it can include planetary scattering and migration (see reference 9). This applies to exoplanetary system Kepler-90 (not presented here) which we found to adhere to TB low with fitting percentage of 72.22 and is estimated to have an age of 0.5 Gyr. The other reason for deviating from the TB law may happen if the system under consideration may be too old, and consequently may suffer from dynamical changes that affect the positions of the planets in the system. This applies to system Kepler 444 presented above which is thought to be 11.2 Gyr old. Similarly, the Kepler-11 system is found to have fitting percentage of 73.24 which is much below the average.

To get an idea about the adherence of our solar system to TB law using the method presented here we note that the overall fitting percentage we get for it is about 85.80%. This is less than the average which we have obtained for the set of 43 systems that we have studied. However, it should be noted that the relative error we get for Neptune is about 27.46% whereas the relative errors for the other 7 planets inside the orbit of Neptune including the Asteroid Belt ranges from 1.0% to 5.26%. If we would include Pluto in these calculations we find that the relative error in its position is about 95.74%. But if we exclude Pluto and do not count it as a planet as it was decided by the IAU in 2006, then we obtain a fitting of 94.89%.

The results presented in our study shows that, assuming high confidence in the TB law applicability and adherence, we can have systematic speculations in respect of diagnosing the status of some pathological exoplanetary system which shows low fitting percentage. Such speculative diagnosis can be tested by observations, including possible errors in observations or calculations. For example, the system Kepler-33 is found to have a fitting percentage of 78.09 which is again much below the average percentage for the set we considered, despite the fact that it is nearly an ideal system harboring 5 planets with the host star being Sun-like star with age of about 4.3 Gyr. This suggest that here we have a pathological case for further studies. Other systems are also analyzed in this respect and could prove that the degree of adherence to TB law might be very useful for analyzing exoplanetary systems.

---